# Towards Spectrally Efficient and Physically Reconfigurable Architectures for Multibeam-Waveform Co-Design in Joint Communication and Sensing

Najme Ebrahimi, *Member, IEEE*, Arun Paidmarri, *Senior Member, IEEE*, Alexandra Gallyas-Sanhueza, *Member, IEEE,* Yuan Ma*, Graduate Student Member, IEEE*, Haoling Li, *Graduate Student Member,*  Basem Abdelaziz Abdelmagid, *Graduate Student*, *IEEE*, Tzu-Yuan Huang, *Member*, *IEEE,* Hua Wang, *Fellow, IEEE*

*Abstract*—Joint Communication and Sensing (JCAS) platforms are emerging as a foundation of next-generation mmWave (MMW) and sub-THz systems, enabling both high-throughput data transfer and angular localization within a shared signal path. This paper investigates multibeam architectures for JCAS that simultaneously optimize waveform shaping and beamforming across the time, frequency, code, and direct analog/ radio frequency (RF) domains. The paper compares Orthogonal Frequency-Division Multiplexing (OFDM), Frequency Modulated Arrays (FMA), Time-Modulated Arrays (TMA), direct RF/MMW modulation, and Code-Division Multiple Access (CDMA)-based systems with respect to spectral efficiency, beam orthogonality, latency, and Angle-of-Arrival (AoA) estimation accuracy. The results highlight architecture-specific trade-offs among beam agility, efficiency, accuracy and resolution, and complexity. It also provides a framework for selecting JCAS front ends optimized for power, latency, inter-beam and multi-user interference, and rapid system reconfiguration.

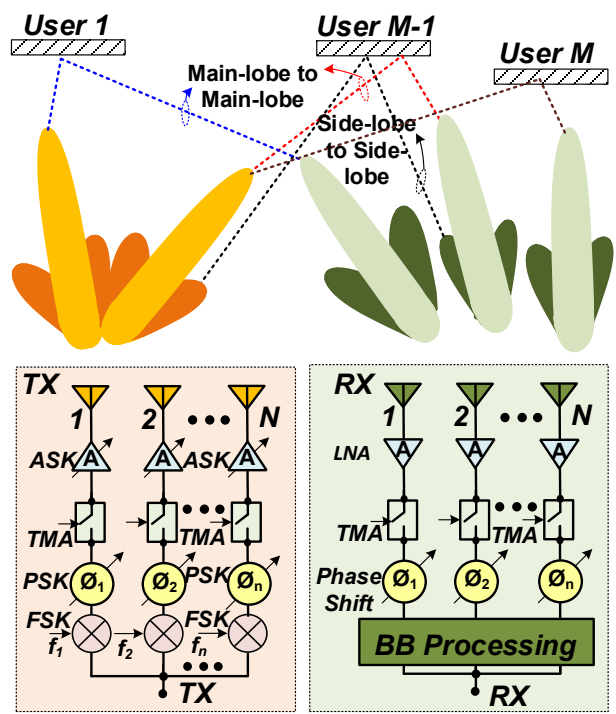

Fig. 1: Frequency, time, code, direct analog/RF array processing for multibeam/users joint communication and sensing.

## I. INTRODUCTION

NEXT Generation of mmWave and sub-THz wireless systems are increasingly expected to support both high-speed communication and accurate situational awareness (sensing) within the same hardware and signal path [1], [2], [3], [4]. Joint Communication and Sensing (JCAS) or Integrated Communication and Sensing  platforms are emerging to meet this need by combining data transmission and spatial sensing into a shared transceiver architecture.  In these systems, the signal needs to carry information, enable spatial awareness via Angle of arrival (AoA) or Direction of Arrival (DoA) estimation, and adapt to dynamic conditions, all through a single, co-designed waveform and beam pattern. Achieving this integration presents a key challenge: waveform shaping and beam shaping are no longer separate design problems. The waveform's spectral, temporal, and modulation properties directly affect the spatial distribution and timing of transmitted energy, while beamforming decisions impact signal bandwidth (BW), latency, and modulation efficiency. These tradeoffs become even more critical at MMW band, where dense antenna arrays, strict power budgets, and fast beam agility demands joint optimization essential.

Phased arrays remain fundamental to JCAS/ISAC, enabling spatial selectivity for both directional communication and angle-aware sensing. However, traditional phased arrays typically steer one beam at a time, limiting their ability to support the concurrency required by next-generation JCAS/ISAC systems. To address this, emerging architecture extends beyond conventional phase control by leveraging time [5], [6], [7], [8], frequency, [9], [10], [12], [13], [14], code, [15-23] or direct RF/analog-domain, [24-38], multiplexing and array-element processing to generate multibeam patterns, interference and inter-beam leakage cancellation and simultaneously extract angular information, Fig. 1.

The frequency-based systems, such as IBM's Eye-Beam, [13], [14], achieve AoA estimation within a single symbol by rapidly switching receive beams across OFDM subcarriers. However, it imposes design tradeoffs including latency, resolution, symbol timing constraints, and peak to average power ratio (PAPR) overhead. The Frequency Modulated Arrays (FMA), [9], introduce frequency shifts and modulations across antennas, encoding spatial directions into waveform arrival time and shape. However, it enables one-shot AoA estimation suitable only for active localization, not passive radar. In addition, it requires element-level multipliers in analog FMA implementations for precise phase and frequency coherency.  Time-modulated arrays (TMA) offer

*IEEE Solid State Circuits Magazine*

another form of spatio-temporal synthesis, steering multiple beams by switching element states with sub-ns timing [7],[8], generating spectral harmonics for multibeam communication and sensing. This demands careful duty-cycle design to suppress inter-beam and inter-harmonics interference, and time modulation frequency should be larger than the modulation bandwidth of the JCAS waveform to prevent aliasing. In addition, direct RF/MMW modulation architectures, including [24-26], [32-36], integrate waveform and beam synthesis directly at RF/MMW frequencies. These systems relax baseband (BB) requirements, especially on analog-to-digital converters (ADC) or digital-to-analog converter (DACs) bit resolutions, offering power-efficient solutions over wide bandwidth modulation. However, achieving high-speed modulation introduces challenges, including direct impedance variation across the array, limited rise/fall time control, and modulation linearity [26], [34]. Finally, CDMA-based multibeam systems, such as [15], construct concurrent beams using orthogonal spreading codes. While this supports fast beam calibration and simultaneous user tracking, it also introduces code leakage, requiring long codes and wideband ADC, proportional to code length. Each of these architectures presents a distinct set of trade-offs in spectral efficiency, sensing accuracy, beam agility, inter-beam and harmonics leakage, and implementation complexity.

In this paper, we aim to present a unified multi-paradigm design framework for JCAS/ISAC architecture, providing a comprehensive analysis of waveform-beamforming co-design for next-generation mmWave JCAS/ISAC systems. Sec. II reviews various multibeam and multiplexing architectures, including OFDM, frequency/time, and code multiplexing as well as direct modulation/multiplexing. Sec. III systematically compares these techniques by examining key theoretical metrics such as spectral efficiency, signal-to-noise ratio (SNR), Cramér–Rao Bound (CRB) for AoA estimation accuracy, latency, performance degradation due to beam misalignment and leakage, and crossbeam orthogonality under multibeam operation. Sec. IV discusses four state-of-the-art (SoA) hardware implementations, emphasizing circuit-level trade-offs. Finally, Sec.V concludes the paper.

## II. ARRAY PROCESSING SYSTEM ARCHITECTURE OVERVIEW FOR COMMUNICATION, SENSING, AND BEAMFORMING

### A. OFDM with Time-slices Beams

Standard OFDM-based approaches, used in IEEE 802.11ad/ay and 5G NR, can overlay sensing functionality on communication waveforms, [13], [14]. IBM's Eye-Beam platform, [14], as shown in Fig. 2(a), demonstrates a digital-domain JCAS/ISAC system that overlays sensing on top of these OFDM waveforms. The approach rapidly switches between predefined receive beams within the duration of a single OFDM symbol, enabling simultaneous data demodulation and AoA estimation, Fig. 2 (a). This is achieved by slicing the symbol in time and assigning each slice to a beam direction. The design trade-off and optimization include fast beam switching, on the order of nanoseconds, and tight alignment with symbol clocks, to concur its sensitivity to timing mismatch and hardware limitations. In practice, the number of usable beams per symbol is limited by switching

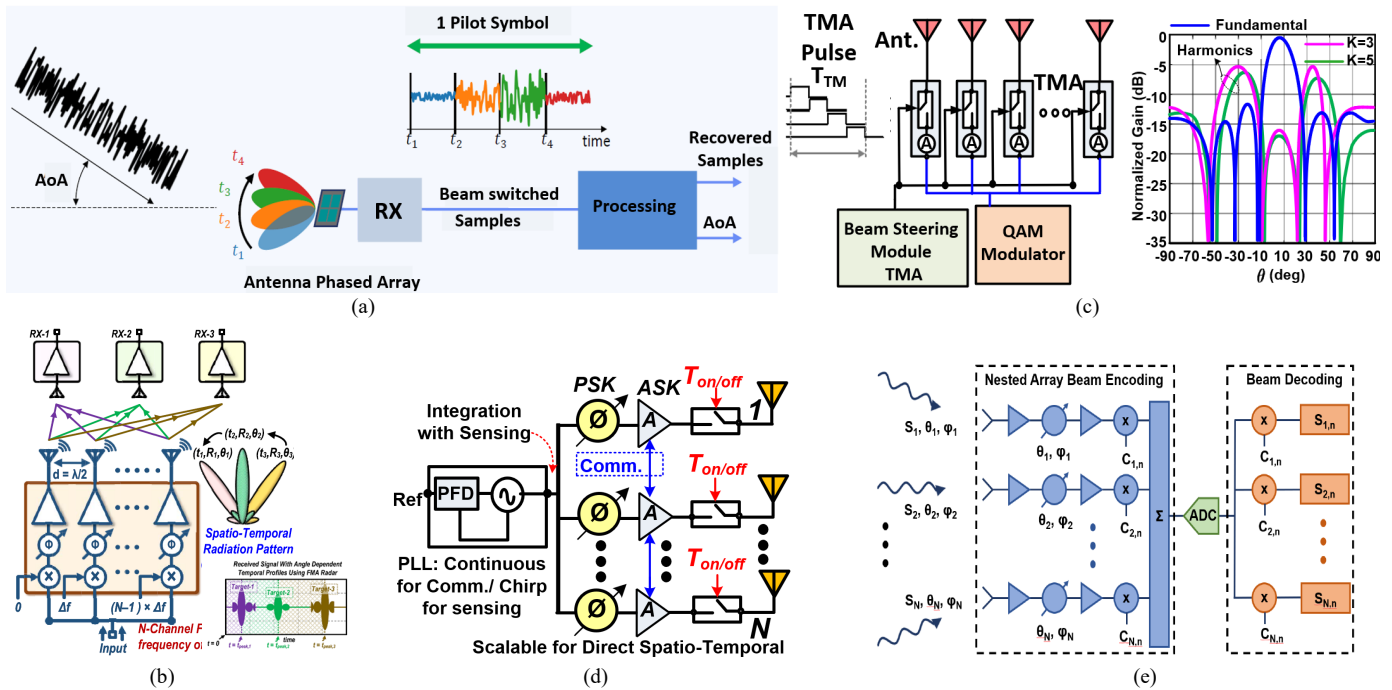

**Fig. 2:** Array processing approaches using frequency, time, code-based multiplexing, and direct RF/analog modulation for JCAS/ISAC, (a) IBM's OFDM-based agile beamforming approach for angle-of-arrival estimation, performing rapid beam switching within a single OFDM pilot symbol duration [13-14], (b) ETH Zurich's Frequency-Modulated Array concept illustrating frequency offsets across antenna elements to achieve angle-dependent temporal radiation patterns [9]. (c) Time-Modulated Array concept, [7], demonstrating separation of fundamental and harmonic beams for concurrent communication and sensing, (d) directional modulation architecture integrating amplitude/phase-shift keying for direct RF spatio-temporal modulation, suitable for scalable high-speed communication/beamforming and radar integration, [24], [29-30] (e) UCSD's CDMA-based nested array utilizing orthogonal codes for multibeam encoding [15].

speed and symbol duration, creating a trade-off between angular resolution and latency.

### B. Frequency Modulated Array

The frequency-modulated array, in [9], introduces a fundamentally different approach to beam shaping and AoA estimation by embedding spatial information directly into the time-frequency structure of the transmitted waveform, as shown in Fig. 2 (b). Rather than switching beams or cycling through codebooks, FMA architectures apply small frequency offsets ($\Delta f$) or chirp variations to each antenna element, creating a time-varying phase gradient across the array. This results in a dynamic far-field interference pattern, where each direction of arrival maps to a unique temporal waveform peak ($t_{peak}$) at the receiver. Consequently, FMA enables one-shot, full-field-of-view AoA estimation between one transmitter array and any number of receivers without requiring beam sweeping, digital calibration, or symbol demodulation. Angular resolution depends on the frequency step size, $\Delta f$, and ADC sampling rate rather than on aperture size or number of elements under ideal conditions. However, the approach requires optimization and design trade-offs on per-element frequency generation and local oscillator (LO) distribution between the TX elements in the same TX array. Another limitation is that it supports angular localization between active TX and RX nodes, however it cannot perform radar-like detection on passive targets.

### C. Time Modulated Array

The TMA system generates multiple directional beams by periodically switching each antenna element in time, introducing spectral components, sidebands as separate beams pointing to other directions, that can be individually steered and modulated, Fig. 2 (c), [7]. Unlike phased arrays that form a single beam through static phase shifts, TMAs encode spatial information into the timing of on/off switching patterns across the array. Each element is pulsed or phase-modulated at a modulation frequency, $f_m$, producing radiation not only at the carrier, $f_c$, but also at harmonics, $f_c \pm n f_m$, as sideband beam, where $n$ is the sideband order. By controlling the phase delays and pulse sequences per element, these sideband beams can be directed toward different angles, enabling simultaneous multibeam operation from a single RF feed. To avoid spectral overlapping and harmonic interference in the main beam, the bandwidth of each beam's signal must remain smaller than $f_m$. This requirement introduces a fundamental trade-off between the achievable number of beams and the available per-beam data rate as well as power consumption and complexity to generate fast time modulation signals. Further, it has been shown that TMA can be combined with traditional static multi-beam array architectures, such as Butler matrix, to achieve "beam multiplication" effects and generate a large number of independent concurrent beams using a small-sized array, particularly useful for MIMO radar applications. In [8], this concept was demonstrated as 20 beam generation using a 4-element array at D-band.

### D. Direct RF/MMW Modulation/Processing for ARRAY

Direct RF modulation architecture offers a fundamentally different pathway for JCAS/ISAC by generating the modulated waveform directly at mmWave or RF carrier frequencies, reducing the complexity on ADC/DACs and baseband up/down conversion. In these systems, each antenna element or subarray is driven by a local modulator that synthesizes both the data waveform and its spatial characteristics. Transmitter architectures such as, [24-38], demonstrate this principle using current-steering DACs and harmonic-rich waveform synthesis to directly produce high-order modulations (e.g., 64 Quadrature Amplitude Modulation (QAM)) at mmWave. Modulation and beamforming are jointly embedded in the RF waveform: phase shifts, amplitude weights, or duty cycles are applied at the element level, Fig. 2(d). This convergence of waveform generation and beam synthesis at the hardware level points toward a new class of mmWave transceivers that are compact, fast, reconfigurable and adaptive.

For JCAS/ISAC applications, this architecture presents two key advantages. First, it enables wide instantaneous bandwidth, since the modulated waveform is generated at RF, chirps or spread-spectrum codes retain their full bandwidth and phase coherence, ideal for radar and high-data-rate links. Second, direct RF synthesis reduces transceiver complexity and supports fast transitions between waveform modes. For example, the same modulator can alternate between radar chirps and QAM data on a per-packet basis, with sub-nanosecond switching latency. Phase and amplitude beamforming can be implemented by using digital control of bit streams, or by modulator-level phase/amplitude injection (e.g. On-off keying), making this architecture naturally compatible with TMA-style spatial coding. However, these advantages introduce critical design trade-offs. Maintaining waveform fidelity across the array requires high-speed switching with minimal distortion, and practical non-idealities, such as rise/fall time imbalance and impedance variation. This can degrade error vector magnitude (EVM) performance, particularly for QAM, sensing accuracy, and restrict the scalability and number of array elements.

### E. CDMA-based Multibeam

The code-domain multi-beamforming approach, developed by [15], introduces simultaneous multibeam transmission using a single RF chain, as shown in Fig. 2(e). In this architecture, spatial multiplexing is achieved by applying orthogonal binary codes (e.g., Walsh/Hadamard or pseudorandom PN sequences) across subsets of array elements. Each desired beam direction corresponds to a unique code; the aggregate radiation pattern forms multiple beams concurrently, with beam separation accomplished by matched code filtering at the receiver.

This approach eliminates the need for time or frequency multiplexing and enables simultaneous multibeam transmission without requiring multiple RF chains or phase shifters. However, it introduces key design trade-offs involving code selection, array symmetry, and analog signal integrity. A critical limitation stems from code leakage: even small mismatches in element amplitude, time misalignment and fading, or RF path mismatches can compromise code orthogonality, resulting in residual crossbeam interference. Moreover, the approach faces scalability challenges, as the number of beams increases, maintaining orthogonality and dynamic range requires exponentially longer codes, increasing both array complexity and calibration burden. Another design trade-off is the mixed-signal interface bandwidth, as the ADC must support *N* times of the bandwidth, where *N* is the CDMA spreading factor (codeword).

III. PERFORMANCE ANALYSIS OF SPECTRAL EFFICIENCY, BEAM ISOLATION, AND AoA ACCURACY

Robust communication, sensing, and beamforming under multibeam and multiuser operation remain challenging in phased array systems and JCAS/ISAC platforms. Achieving high throughput, low beam alignment latency, and accurate angle-of-arrival estimation is fundamentally constrained by hardware and RF nonidealities. Signal integrity is degraded by interference, leakage, amplitude and phase mismatches, as well as timing misalignments, synchronization drift, and multipath-induced delays. These impairments disrupt beam orthogonality, reduce angular resolution, and elevate sidelobe levels, resulting in inter-beam interference that limits latency, user scalability, and spectral efficiency.

To systematically evaluate these architectures, we assess four key performance dimensions: spectral efficiency, inter-beam interference, angle-of-arrival estimation accuracy, and latency based on fundamental theory. Each metric reflects a fundamental trade-off between signal processing complexity, hardware precision, and multiplexing strategy, [39-43]. The analysis is organized in the following subsections, with a summary of performance benchmarks outlined in Table I.

A. *Spectral Efficiency*

Spectral efficiency ($\eta$), defined as the throughput per unit bandwidth (bits/s/Hz), is critical for evaluating multibeam/multiuser JCAS/ISAC architectures, which can be expressed as (1):

$$\text{Spectral Efficiency } (\eta) \propto J \log_2\left(1 + \frac{SNR}{IS(\tau,f)+1}\right) \quad (1)$$

In (1), $J$ is number of beam/users, SNR is signal to noise ration and $IS(\tau,f)$ defines as isolation between beams/user over frequency, $f$, and time drift, $\tau$. In Time-Division Multiple Access (TDMA), TMA, FMA, and OFDM approaches, the spectral efficiency per beam typically decreases as the number of beams increases, due to limited bandwidth resources shared among beams. In contrast, CDMA systems enable all beams to occupy the entire bandwidth simultaneously by employing orthogonal spreading sequences (e.g., Walsh-Hadamard codes). Similarly, in TMA arrays, precise time modulations are required to ensure the quality of the multi-beams and inter-beam isolations. Therefore, practical imperfections, such as timing jitter and synchronization drift, introduce cross-correlation between codes, causing interference that reduces the effective isolation and thus spectral efficiency. Direct RF modulation with hybrid beamforming achieves enhanced spectral efficiency through spatial diversity. The performance of direct RF modulation architectures critically depends on interference suppression, represented by $IS(\tau,f)$, which is directly influenced by signal bandwidth. Consequently, spectral efficiency in direct RF modulation scales favorably with the number of simultaneous beams, emphasizing the importance of mitigating inter-beam leakage and interference. Table I summarizes these trade-offs, highlighting the impact of multiplexing methods and practical limitations on spectral efficiency in various multibeam/multiuser architectures.

B. *Multibeam/Multiplexing and Inter-beam Leakage*

Efficient operation of multibeam JCAS/ISAC systems demands robust isolation between simultaneously transmitted or received beams. However, practical challenges such as timing errors, synchronization drifts, and multipath propagation often degrade beam orthogonality, increasing inter-beam interference and leakages. The CDMA architecture relies on orthogonal spreading sequences, typically Walsh-Hadamard codes, to separate concurrent beams such as the work in [15]. Similarly, RF-based signal processing methods using CDMA-based code-select and code-reject *N*-path filters and full-duplex transceivers have been developed in [21-23]. However, these code-based systems fundamentally affect the system performance through isolation among multi-beams/users, or transmitter (TX)/Receiver (RX) pairs, depend heavily on timing alignment and fading.

Table I: Comparison Between Various Array Processing for Multiplexing/Multibeam JCAS/ISAC

| Metric/Technique | TMA | OFDM(FMA) | CDMA(Walsh) | Direct Modulation |
|---|---|---|---|---|
| Spectral Efficiency($\eta$) | $\frac{1}{J} \log_2(1 + SNR)$ | $\frac{1}{J} \log_2(1 + SNR)$ | $\log_2\left(1 + \frac{SNR}{1 + \frac{J-1}{N}}\right)$ | $J \log_2\left(1 + \frac{SNR}{IS(\tau,f) + 1}\right)$ |
| Beam Isolation(IRR) | Moderate | Moderate | $10 \log(1/N)$ | Moderate |
| Fading Robustness | Time Diversity | Frequency Diversity | Code Diversity | Time Diversity |
| AoA Estimation Error (CRB) | $\frac{CRB_\theta}{2 SNR_{eff} \cdot N_{tot}/J}$ | $\frac{CRB_\theta}{2 SNR_{eff} \cdot N_{sym} \cdot N_{sub}}$ | $\frac{CRB_\theta}{2(SNR_{chip} \cdot L)}$ | $\frac{CRB_\theta}{2 \cdot SNR_{eff}}$ |
| Latency | $U. T_{sequential}$ | $U. T_{symbol}$ | $T_{code}$ | $U. T_{sequential}$ |

*N*: number of codeword, *M* number of antenna, J number of multi beam/users, $CRB_\theta = 1/(2\pi \frac{D}{\lambda} \cos\theta)^2$, *U* product of the number of simultaneous beams, and *T* symbol/code/sequential duration, $N_{sub}$ is the number of OFDM subcarriers used, $N_{sym}$ the number of OFDM symbols. For OFDM, time slicing reduces the effective matched filter integration time, and $SNR_{eff}$, leading to an SNR per beam reduction proportional to the slice duration ($\Delta t$).

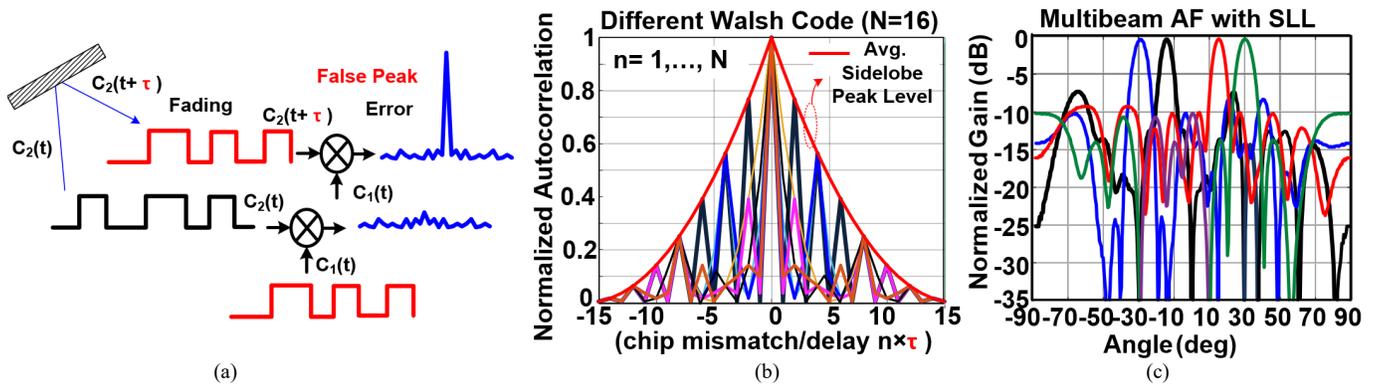

Fig. 3: (a) Conceptual block diagram illustrating the fading-induced cross-correlation effect in orthogonal-code CDMA systems (e.g., Walsh and Hadamard codes), (b) normalized autocorrelation of various N=16 Walsh codes under different chip mismatch or fading delays, (c) multibeam array factor (AF) demonstrating sidelobe interference levels resulting from code multipath delay and fading effects (demonstrating SLL near -7dB).

To evaluate the inter-beam leakage and the impact of timing offsets, the cross-correlation between codes can be analyzed and utilized as a performance metric [44], [45]. The normalized cross-correlation between two orthogonal codes, $\rho_{code}(\tau)$, subjected to a timing offset $\tau$ can be modeled as:

$$\rho_{code}(\tau) = \frac{1}{N}\sum_{n=0}^{N-1} c_i(nT_c) \cdot c_j(nT_c + \tau) \quad (2)$$

Where $N$ is the length of code word, $c_i$ and $c_j$ are the two orthogonal codes and $T_c$ is the duty cycle. Experimental data indicates that a single-chip delay in a 16-chip Walsh-coded system result is $1/N$ cross-correlation, considerably lowering beam isolation under multipath conditions. This results in cross-correlation sidelobes reaching levels only 1 to 2 dB below the peak value (equivalent to a normalized cross-correlation value of approximately 0.8) as shown in Fig. 3(a) and (b). Consequently, substantial inter-beam leakage occurs, raising the normalized side-lobe level (SSL) to nearly -7 dB, as shown in Fig. 3(c). Therefore, the inherent sensitivity of Walsh and CDMA codes to timing misalignment and multipath dispersion significantly restricts their suitability for scalable, low-latency, and robust high-throughput multibeam JSAC.

On the other hand, FMA methods achieve beam isolation through frequency increments across antenna elements, defined as $f_n = f_c + n\Delta f$. In frequency-modulated arrays, signals transmitted at incrementally shifted frequencies coherently combine to form distinct spatial beam patterns. This interference can be quantitatively evaluated through the peak sidelobe ratio, making beam isolation dependent on frequency coherence and synchronization misalignments. On the other hand, TMA arrays form multiple beams by rapidly switching antenna elements, generating harmonic beams at multiples of the switching frequency, $f_m$. To ensure sufficient spectral isolation and prevent sideband overlaps, the signal bandwidth $B$ must satisfy $B \leq (f_m/2)$. Similarly, direct RF/mmWave modulation systems can achieve beam isolation through spatial coding and analog beamforming techniques. However, the time and switching based approaches are also sensitive to hardware imperfections such as finite rise/fall times, duty-cycle mismatches, load modulation effect, bandwidth induced nonlinearities and noise. These hardware-induced imperfections increase sidelobe levels and degrade spatial isolation between beams.

### C. Multibeam AoA Accuracy and CRB

Another key sensing metric in JCAS/ISAC systems is the angular precision of angle of arrival estimation. The fundamental limit is given by the Cramér–Rao Bound (CRB), which defines the lowest achievable variance of an unbiased AoA estimator [43]. For an array of aperture $D$, wavelength $\lambda$, and effective *SNR*, $SNR_{eff}$, the CRB scales as [42], [43]:

$$\text{Var}(\hat{\theta}) \geq \frac{1}{2\, SNR_{eff} \cdot J \cdot \left(2\pi \frac{D}{\lambda}\cos\theta\right)^2} \quad (3)$$

The bound improves with higher *SNR*, wider aperture, and multiple independent measurements, $J$, typically obtained through bandwidth, time diversity, or chirp structure.

For OFDM-based architectures, such as IBM's Eye-Beam platform, AoA estimation is performed by coherently combining multiple subcarriers across OFDM symbols. This results in an effective $SNR_{eff}$ of $SNR_{sub} \times N_{sub} \times N_{sym}$, where $N_{sub}$ is the number of OFDM subcarriers used, $N_{sym}$ the number of OFDM symbols integrated, and $SNR_{sub}$ is the per-subcarrier SNR. The increased effective SNR directly reduces estimation error, achieving finer AoA accuracy for this multiplexing approach. On the other hand, FMA array enhances AoA estimation by introducing incremental frequency offsets across antenna elements, leveraging frequency diversity for improved angular resolution. In theory, if both OFDM and FMA systems use identical bandwidth and *SNR*, their fundamental AoA estimation variance (CRB) would be comparable. Both methods effectively enhance AoA accuracy through increased bandwidth and effective *SNR*; OFDM via multiple subcarriers and FMA through multiple frequency offset tones. The frequency offsets in FMA can be derived from the LO signal by frequency dividing to ensure its quality. Additional multipliers are needed in each element for analog FMA. On the other hand, it is conceivable that the element-level frequency offsets of FMA can be implemented in the backend processing of digital beamforming arrays without hardware multipliers. Similarly, the AoA estimation accuracy in TMA and direct RF/MMW modulation systems can similarly benefit significantly from the multiple harmonic generation, acting as frequency diversity and effectively reducing AoA estimation variance. In other words, employing multiple harmonic beams increases the effective number of independent angular measurements, thereby reducing the overall variance of AoA estimates. Nevertheless, practical challenges as mentioned in Sec. III.A exist, such as finite switching speeds, rise and fall times, duty-cycle imbalances, and timing jitters, which can impose limitations and somewhat increase the variance above theoretical predictions. On the other hand, the CDMA-based arrays leverage orthogonal spreading sequences for simultaneous multibeam transmission, [15]. The CRB for AoA estimation in CDMA depends on the code length, $N$, chip rate, $f_{chip}$, and effective code isolation. Specifically, the AoA estimation variance can be approximated as:

$$\text{Var}(\hat{\theta})_{CDMA} \approx \frac{1}{2\,(SNR_{chip} \cdot N)\left(2\pi \frac{D}{\lambda} \cos\theta\right)^2} \quad (4)$$

The enhanced SNR and integration gain provided by the code-length significantly enhances robustness against interference, allowing CDMA-based arrays to approach the theoretical CRB of digital arrays under ideal conditions. However, practical timing misalignment and fading errors degrade code isolation, Fig. 3 (b), thereby deteriorating AoA estimation accuracy.

In summary, CDMA-based arrays achieve AoA accuracy close to the CRB only when near-ideal timing synchronization is maintained. FMAs rely on stable frequency offsets and precise phase coherence, and additional element-level multipliers in analog FMA implementations. TMAs and direct modulation-based arrays require accurate modulation timing and precise switching control, making hardware precision critical. Therefore, each method presents unique trade-offs among complexity, accuracy, and practicality. Hybrid approaches or combining complementary techniques are often necessary to practically approach theoretical CRB limits. Therefore, employing hybrid calibration and optimization methods can effectively mitigate synchronization errors, frequency instability, and hardware imperfections, significantly improving AoA estimation accuracy under realistic operational conditions.

### D. Latency and Beam Agility

Latency and beam agility are critical metrics that directly impact the performance of multibeam JCAS/ISAC systems, particularly in dynamic environments requiring rapid tracking and precise sensing. OFDM-based systems, such as IBM's Eye-Beam, update beam directions sequentially at the OFDM symbol rate, resulting in a latency directly proportional to the product of the number of simultaneous beams, $U$, and symbol duration, $T_{sym}$. This linear scaling implies that increasing the number of beams proportionally increases the latency. Nevertheless, Eye-Beam achieves sub-10 ns control accuracy, significantly mitigating latency in practical implementations. FMA array, by encoding spatial information through incremental frequency offsets across antenna elements, enable single-shot parallel sensing across multiple angles without sequential scanning. Therefore, FMA latency, in theory, is constant and "one-shot". Although this technique can deliver rapid AoA estimation, controlling frequency offsets or configuring beams introduces practical latency constraints due to practical hardware limitations. TMA arrays offer latency related to simultaneous multiple beams generations through harmonic modulation. Since all harmonic beams are produced within a single modulation cycle, beam updates occur in parallel. This enables a low latency system, limited primarily by the switching frequency and waveform integrity. In contrast, direct RF/MMW beamforming relies on sequential beam steering, where each beam is updated one at a time using tunable phase/amplitude shift keying. As a result, latency is related to the number of beams and the dwell time per beam.

Consequently, achieving optimal beam agility in JCAS/ISAC systems typically involves combining multiple multiplexing methods or adopting hybrid approaches, matching the specific application requirements and practical hardware constraints.

## IV. ARRAY PROCESSING AND CIRCUIT-LEVEL TRADE-OFFS

This section presents states of the arts demonstrations for the four multibeam waveform co-design architectures, from agile beamforming using OFDM (IBM), [13], [14], to multibeam CDMA coded array (UCSD), [15], frequency/time-modulated arrays (ETH), [7], [9], and direct RF/mmWave array for modulation and reconfigurable beamforming, (NU), [24-26]. Each implementation highlights key circuit-level strategies while addressing dominant nonidealities, such as finite switch rise/fall times, transition asymmetry, timing jitter, LO phase noise, and inter-element mismatches, that directly impact the effective SINR/SNR and ultimately govern system-level performance, as discussed in the previous sections:

$$\text{SNR}_{\text{eff}} \approx \frac{P_0\, G_{array}}{kT_0 B F_{sys} + \sum I_{\text{spur,leak}} + \sum I_{\text{ICI,ISI}} + N_Q} \quad (5)$$

Where, $F_{sys}$ captures modulator losses and associated bandwidth, $B$, and noise figure (NF) due to RF impairments, $\sum I_{\text{spur,leak}}$ includes leakage, sideband emissions, and code mismatch terms (e.g., $|\rho_{code}|^2$ in coded based system), $\sum I_{\text{ICI,ISI}}$ accounts for timing jitter, phase noise, and mid-symbol beam-switching artifacts, $P_0\, G_{array}$ received or transmitted power of the array, and $N_Q$ represents quantization noise, critical in shared-path ADC/DAC receivers. These terms highlight how various circuit nonidealities dominate performance in different waveform co-design strategies, that will be elaborated and discussed in next subsections *A* to E.

### A. Coded-based Multiplexing/ Array

In coded modulated or code-multiplexed systems, such as UCSD CDMA-based arrays [15], Fig. 4(a), or code-domain IF/LO multiplexing of MIMO arrays in [18], [19], Fig. 4(b), each antenna path is modulated by a high-speed orthogonal code (e.g., ±1 Walsh sequence). The resulting coded signals are summed and routed through a shared RF/baseband/ADC chain, as shown in Fig. 4 (a) and (b). Digital matched filters (DMFs) are then used to recover each stream (channel/element). The code modulation can be implemented using a code-modulating cascode low-noise amplifier (CM-LNA), as shown in Fig. 4(c), [18], or by LO polarity switching (±LO for BPSK), shown in Fig. 4(d), [19], or at intermediate frequency (IF) before summation. These methods emulate ±1 gain modulation to apply orthogonal Walsh codes.

The code modulators, shown in Fig. 4(c) and (d), are implemented using FET or bipolar switches, which can introduce excess noise due to finite rise/fall time and overlap conduction (Fig. 4). However, because chip rates are well below RF frequencies, this switching noise is upconverted and can be filtered by the resonant load. Measured CM-LNAs show minimal noise figure (NF) penalty at moderate chip rates in [18]. Therefore, proper modulator design with fast, symmetric transitions is essential to minimize spectral splatter and maintain fidelity in coded arrays. In addition, these architectures are susceptible to bandwidth expansion and linearity constraints. Coding spreads the signal according to a code length or processing gain $G$, increasing the required bandwidth for all post-modulation blocks. For example, for $G = 8$, the low pass filter (LPF), variable gain amplifier (VGA), and ADC must support 8× of the baseband bandwidth. Consequently, using quasi-orthogonal codes ($\rho_{code} < 0.6$) can reduce bandwidth but introduces variable leakage, reducing isolation as shown in Fig. 4(e). The design tradeoff lies in balancing system isolation with analog/RF bandwidth and linearity. Furthermore, summing $N$ coded signals increases the ADC input power by ~10log10(N), limiting dynamic range and increasing quantization noise, $N_Q$ in (5). To mitigate this, per-path VGAs are typically used to equalize signal levels. In larger arrays, gain and phase mismatches across elements distort the effective code vectors such as its channel-to-channel isolations and normalized gain as shown in Fig. 4(e). In a 1024-element Ku-band nested subarray demonstrated at UCSD [15], beam isolation was limited to a sidelobe level of 20 dB due to phase/gain path errors. Maintaining code orthogonality in such cases requires array calibration or digital compensation using the reference encoding matrix.

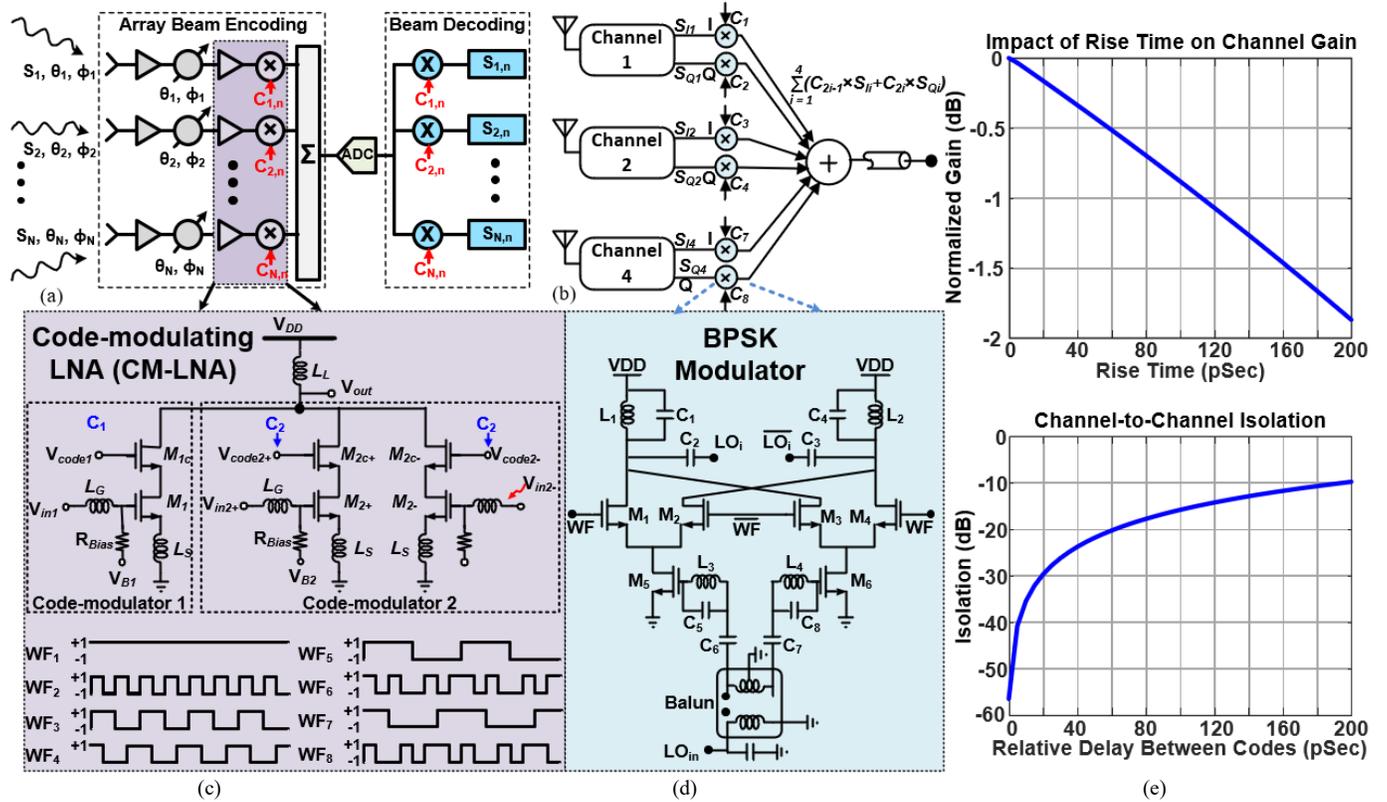

**Fig. 4:** Conceptual diagram code-domain multiplexing array and the circuit block diagram, (a) CDMA nested array in [15] and (b) Code-domain multiplexing (CDM) receiver for IF path sharing in [19], (c) Code-modulating LNA (CM-LNA), [18], with associated Walsh coded waveforms, (d) BPSK modulator for multiplying LO with Walsh codes, [19], and (e) Top: Impact of rise time for the WF codes on the channel gain, bottom: Impact of relative delay between codes on channel-to-channel isolation in [19].

In summary, when timing, code selection, and analog hardware are carefully co-designed, CDMA/code-modulated/multiplexed receivers can approach the performance of conventional architectures with significantly lower hardware overhead. The dominant circuit-level impairments, including timing misalignment, switching distortion, and ADC quantization, can be mitigated by on-chip calibration to keep SNR degradation small, consistent with the system-level analysis presented in Sec. II and III.

### B. Frequency Modulated Array

The FMA transmitters apply slightly frequency offsets ($\Delta f$) across elements to encode angle information in the received signal's time-domain pattern as shown in Fig. 5 (a). The ETH's FMA/TMA implementation, [7], [8], [9] enables real-time *AoA* sensing by introducing either frequency offsets or periodic switching across array elements, Fig. 5 and 6. In the FMA system, MHz-scale frequency offsets are applied between adjacent antennas, causing the relative phase across elements to vary in time. This results in a unique temporal pattern at the receiver, where the timing of amplitude peaks, or equivalently, the beat phase, depends on the signal's angle of arrival. However, this scheme is highly sensitive to inter-channel timing and phase alignment. Any skew in the $\Delta f$ modulation timing between antenna channels can affect the intended *sinc*-like temporal radiation pattern. In terms of the circuit implementation, instead of on/off switching, FMA uses analog multipliers or mixers at IF to introduce the $\Delta f$ offsets, Fig. 5 (b). The core consists of a differential input pair that modulates transconductance, followed by a double-balanced switching quad that acts as a mixer, Fig. 5(c). Analog multipliers introduce several circuit-level nonidealities that can degrade angular accuracy and signal quality, *SINR* in (5). Nonlinear transfer characteristics can generate higher-order harmonics (e.g., $f_{RF} \pm 3\Delta f$), which interfere with timing detection and angle-of-arrival estimation. These spurs are further deteriorated by gain compression, mismatch between differential paths, and limited bandwidth. To minimize these effects, ETH's FMA transmitters employ class-*A* biased multipliers with symmetric differential layouts, Fig. 5(c), and with on-chip filtering, and symmetric LO and bias routing. These design measures suppress spurs below –28.5 dBc and help preserve spectral purity and timing resolution. Simulations and measurements show that delay mismatches of 5-100 ns can introduce angular errors ranging from 0.64° to 17.4° (Fig. 5 (e) and (f)). Phase-aligned LO paths and matched lengths routings are used to constrain timing skew within a few nanoseconds. As shown in Fig. 5(e) and (f), real multipliers induce modest angle mismatch (~±2.5°), that can increase errors to ~15° with higher spur levels (–10 dB). In summary, the circuit-level limitations in FMA systems, including delay mismatch, modulation spur content, and sampling resolution, can be minimized through careful LO distribution, analog linearity, and backend ADC design. As demonstrated in ETH FMA system, a 28 GHz 4-element transmitter chip, fabricated, supports dual-mode operation: (a) standard phased array transmission with quadrature phase shift keying (QPSK)/64-QAM modulation, and (b) FMA sensing via frequency offsets. The resulting far-field pattern effectively scans the angular space without physical beam steering. Experiments using a PCB-based patch array demonstrated full ±60° field-of-view coverage and ~2° resolution. In sensing mode, the same FMA transmitter distinguished objects spaced 1° apart, significantly outperforming conventional arrays of similar size.

### C. Time Modulated Array

In contrast, TMA-based MIMO receivers, such as ETH's TMA in [7], use high-speed RF switches periodically gate each antenna element following a synchronized time-sequence, enabling spatial-to-spectral mapping of incident beams onto distinct harmonic tones. Each element is typically connected to a tail-switched LNA stage, [7], Fig. 6(a), which supports fast ON/OFF modulation with optimized added insertion loss. The switching is driven by digital clock signals, and the resulting modulation folds spatial angle information into frequency components centered at $f_c \pm m f_m$, where $f_m = 1/T_m$ is the modulation frequency. Non-instantaneous switching edges (e.g., >10-20 ps) leak energy into unintended harmonics, raising the spectral power for adjacent beams. Asymmetric switch transitions or clock delays, Fig. 6(a), distort sideband symmetry, degrade main-lobe gain, and reduce sidelobe suppression. These distortions limit beam isolation and directly affect *SINR in* (5), especially for multi-beam MIMO reception. While the use of tail-current switching LNA circuitry minimizes RX insertion loss at lower mmWave frequencies, it significantly degrades OFF-state isolation, dropping below 40 dB at 30 GHz and below 10 dB above 100 GHz, which results in leakage from OFF paths into active beams, thereby increasing inter-beam interference. Therefore, stringent filter and resonant loading are required to meet isolation and leakage requirements. Furthermore, timing jitter, delay or mismatch between switching clocks across elements manifests as effective phase noise on the time-modulation frequency $f_m$, breaking harmonic orthogonality. Simulations in Fig. 6 (b) show that a 0.1 ns skew or 8° phase mismatch between two channels can reduce crossbeam isolation from ideal (>40 dB) to ~25-27 dB at 30 GHz. The performance at higher mmWave frequencies (>100 GHz) will be discussed in Sec. IV. *E*.

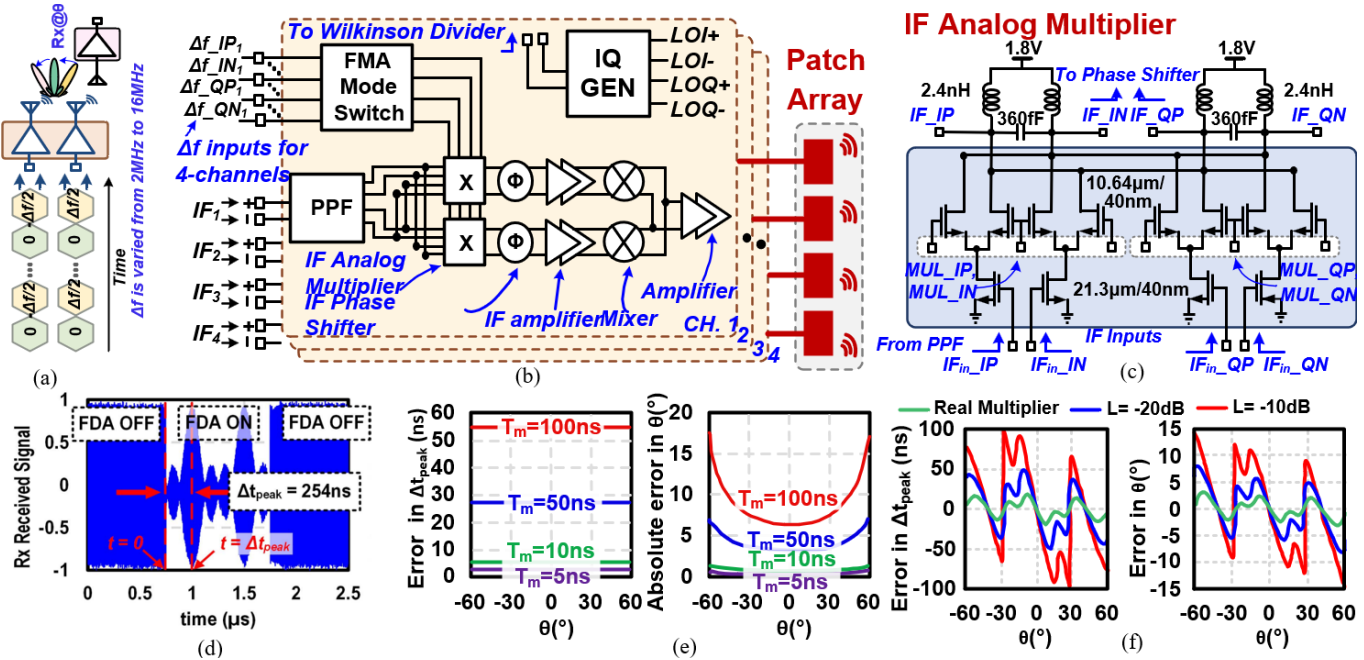

**Fig. 5:** Frequency-modulated FMA array, [9] (a) Input timing sequence for applying MHz-scale frequency offsets between adjacent antennas, causing the relative phase across elements to vary in time, (b) Circuit block diagram for the mm-Wave multi-channel FMA Tx including analog multipliers, (c) Schematics of analog multiplier for FMA operation and mode-switch for multimode operation of Tx, (d) An example of measured transient waveforms at the Rx using the transmitted of FMA signals sequence in (a), (e) Effect of delay mismatch in *Δf* signals versus time delay and different angle (f) Effect of analog multiplier additional *Δf* harmonics due to the path errors and time delay.

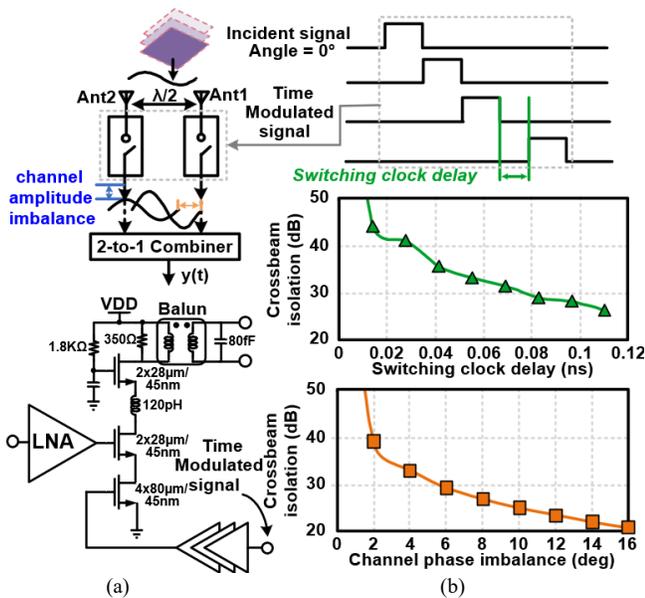

**Fig. 6:** Time modulated array with spectral-spatial mapping in [7] and time sequence, as well as switched cascode LNA for on-off keying, (b) Simulated cross beam isolation of two element TMA under phase mismatch and TM switching clock delay.

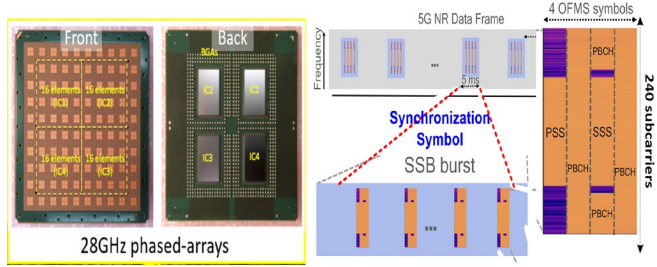

**Fig. 7:** IBM's platform consisting of a 64-element phased array, software-defined radio (SDR) for I/Q data acquisition. The illustration shows synchronization signal blocks transmitted in bursts from a 5G NR gNodeB, detailing the operational principle and digital processing method used for AoA estimation based on digitized signals collected by beam sweeping with 4 beams, [13], [14].

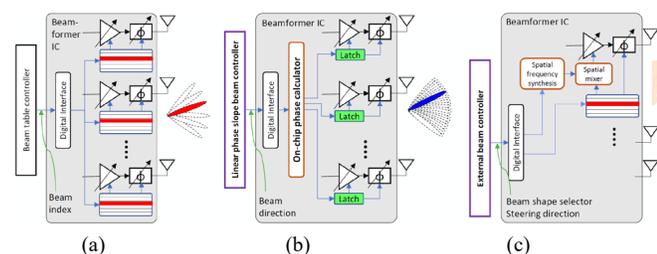

**Fig. 8:** Digital control architectures for fast beam switching for OFDM-based phased arrays, (a) Beam-table (SRAM) method: a beam-table approach where SRAMs store phase/gain settings for each beam, (b) On-chip beam calculator for phase/amp calibration, (c) Hybrid arbitrary-shape steering: fast steering of arbitrary beam that combines the advantages of the other two approaches in (a) and (b).

### D. OFDM-Based Agile Beamforming for JCAS

IBM's platform [13] presents an integrated JCAS/ISAC architecture using a 64-element software-defined phased array operating in the 28 GHz band, [13, 14], Fig. 7. The system utilizes agile, sub-symbol beam switching to decode OFDM symbols while simultaneously estimating the AoA. The system synchronizes precisely (<10 ns timing synchronization) the beam switching and the baseband I/Q samples. By rapidly switching among multiple receiver beams within a single OFDM symbol (up to 16 beams withing an 8.33μs symbol), the received waveform experiences a characteristic time-domain modulation unique to the AoA, enabling accurate angle estimation, Fig. 7.

Specifically, the system slices each OFDM symbol, such as the Primary Synchronization Signal (PSS), one of the symbols within the Synchronization Signal Block (SSB) used in 5G, into distinct time intervals corresponding to different beam directions. As the receiver cycles through beams, the ideal waveform of the known PSS symbol is modulated by beam pattern gains, introducing distinct time-domain features. A correlation-based processor compares the multiple received modulated waveform against all possible true AoA values and time-shifted known PSS waveforms, accurately determining the ID of the specific received PSS signal, the AoA and synchronization timing. Experimental results indicate an AoA estimation accuracy within ±3° at SNR levels above 12 dB and below ±5° even at 7 dB SNR, provided six beams are used [13].

*1) Digital control architectures for fast beam switching in phased arrays:* The key to implementing JCAS using time-switching over OFDM symbols is the ability to: (a) achieve reliable and accurate beam shapes, (b) access to a large set of beams that meets the application needs, (c) fast switching among the beams. The phased arrays used in IBM's examples achieve reliable and accurate beam shapes through the use of phase-independent gain control in the VGAs, gain independent phase control in the phase shifters and antennas with uniform and identical radiation patterns [13,14]. The digital architecture to control the beams determines the latency of beam switching and the number of available beams. Fig. 8 shows state of the art digital architecture for beam control. Fig. 8(a) shows a beam-table approach where SRAMs store phase/gain settings for each beam for each front-end in the array. The beams can be switched by simply changing the beam index which can be broadcast to all the front-ends. The beams can be switched within a latency of 240ns with the beam switching over the air within 4ns. However, the number of beams available for fast beam switching is limited by on-chip memory. In contrast, Fig. 8(b) shows an approach of an on-chip calculator that computes the phases required for each front-end using the location of each antenna and the beam direction (represented in linear phase slopes along the x and y-axes of the array) [13, 14]. While only linear-phase slope beams can be generated, this architecture provides fast beam switching among > 30, 000 unique beams within a 200ns latency. This is because only the phase slopes are broadcast to all the front-ends. Fig. 8(c) shows a recent approach for fast steering of arbitrary beam that combines the advantages of the other two approaches. Each front-end stores the phase and gain parameters for arbitrary beam shapes in memory. A spatial mixer applies a linear phase slope (computed by a spatial frequency synthesizer) to steer the arbitrary beam shape in space. This takes advantage of the fact that the beam shape in space is related to the applied phase and gain per front-end through the Fourier transform. An index to select the arbitrary beam shape and the linear phase slopes for steering can all be broadcast to all front ends, enabling fast beam switching (200ns latency). In summary, IBM JCAS system for AoA switches among as many as 16 beams within one 5G-NR OFDM symbol. While the demonstration uses a beam-table approach, the beam calculator approaches could be used to improve operation.

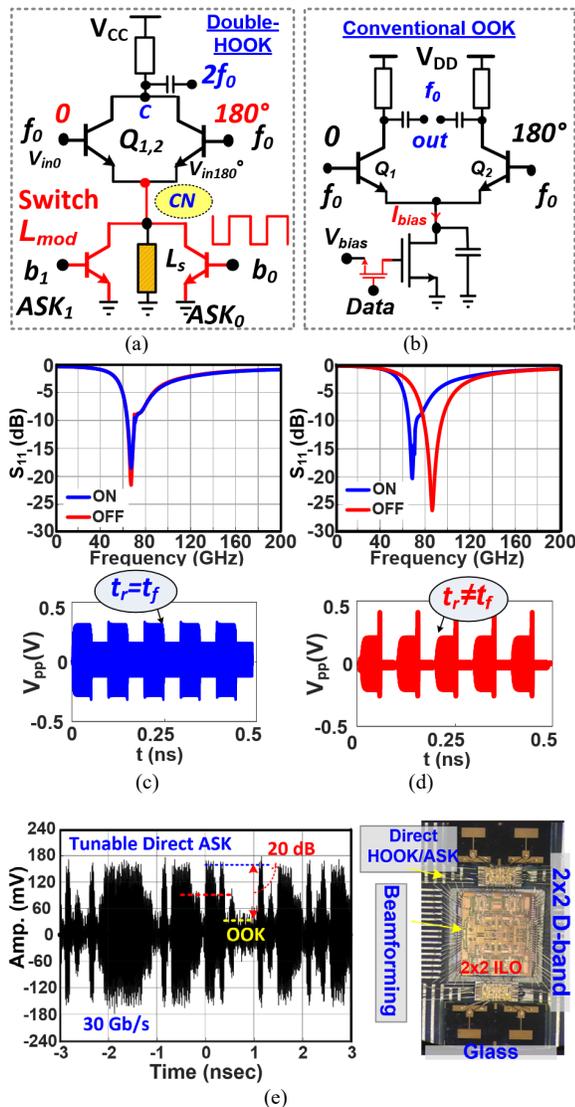

**Fig. 9:** Direct modulation circuit architectures: (a) Active common-node (CN) Double-HOOK modulator demonstrating constant-impedance switching [24-28] (b) Conventional differential amplifier OOK modulator under bias switching [31-34]; (c) reflection coefficient ($S_{11}$) versus frequency for Double HOOK modulator under switching/modulation with transient waveforms with equal rise and fall time; (d) Reflection coefficient ($S_{11}$) versus frequency for conventional modulator under switching/modulation with transient waveforms with unbalanced rise and fall time; (e) The 3-level ASK modulation and 2x2 array scalability of Double HOOK with beamformer and antenna array.

### E. Direct Amplitude Shift Keying Modulation Array

At MMW frequencies, direct modulation and demodulation of complex waveforms such as OOK, ASK, APSK, and QAM require high-speed amplitude and phase keying tightly integrated with pulsed waveform generation at both the transmitter and receiver, [24]–[38]. Implementing these architectures in silicon presents multiple challenges. First, achieving high-frequency modulation and demodulation demands wide instantaneous bandwidth and fast, precise switching, pushing the limits of transistor speed and signal fidelity. Second, maintaining this bandwidth without signal distortion requires constant-impedance, symmetric switching behavior to suppress ringing, overshoot, and waveform distortion. Third, finite transition times and asymmetry between rise and fall edges introduce spectral leakage, resulting in unequal sidebands and degraded spectral purity. These impairments increase BER and necessitate higher SNR to maintain link quality. For array-based systems such as JCAS or time-modulated arrays, additional challenges arise from VSWR mismatches between the power amplifier and antenna array, or between LO drivers and switching networks, which lead to amplitude and phase mismatches across elements. Such mismatches degrade coherent beamforming and reduce system scalability. Together, these limitations highlight the need for new circuit-level architectures that directly embed high-speed, wideband modulation into the RF path while minimizing impedance variation, power consumption, and calibration overhead.

To overcome the limitations of conventional mmWave direct modulation architectures, namely, impedance instability, asymmetric rise/fall times, and spectral leakage, researchers at Northeastern University (NU) introduced a harmonic on-off keying (HOOK) modulator based on a "common-node" (CN) topology, [24]–[28]. In this approach, switch modulators are placed at the virtual ground (common node) of a differential push-push class-*B* mode frequency doubler as shown in Fig. 9(a). In even-mode operation, the doubler employs emitter degeneration with the switch impedance located at the CN, shunted by a short transmission stub. The switches operate in the triode region and are tuned to resonate at 140 GHz. Fabricated in 90 nm SiGe BiCMOS, this architecture leverages bipolar transistors with significantly higher ON/OFF resistance ratios than CMOS-based designs, at above 100 GHz MMW band. The circuit achieves >10 dB isolation compared to ~6 dB reported in earlier CMOS CN-OOK work [25], Fig. 9 (b). This enhanced switching behavior supports a double-switch configuration with different transistor sizes, enabling 3-level ASK modulation with ~5 dB spacing between amplitude levels, Fig. 9 (e). The CN placement ensures nearly constant input impedance during switching transitions (<1% variation), minimizing load transients and avoiding overshoot or ringing, as shown in Fig. 9 (c). HOOK prototypes demonstrate fast, matched rise/fall times (~10% settling time) and support modulation speeds up to 10 GHz. In contrast, conventional amplifier-biased OOK/ASK modulators exhibit large impedance swings and ~40% asymmetry, leading to waveform distortion and degraded spectral purity, Fig. 9(d). The CN-based modulation also suppresses even-order distortion and results in cleaner transients with higher throughput. This architecture is particularly well-suited for scalable array integration, where edge symmetry and impedance stability are essential for minimizing inter-element mismatch and out-of-band emissions. The recently developed architecture integrates a 2×2 glass-based antenna array operating at D-band, featuring HOOK modulators centered at 140 GHz, shown in Fig. 9(e) for scalability illustration. The system utilizes an injection-locked 2×2 LO beamformer at 70 GHz based on the design from [46], [47] that can be heterogeneously integrated with array of HOOKs for time, amplitude and phase modulation/switching with minimum loading on the beamformer IC. This practical implementation demonstrates significant improvements in spectral efficiency and system scalability, for next generation of coded and multiplexed array, desired for JCAS.

## V. Conclusion

This paper systematically analyzes and compares multibeam waveform co-design architectures for joint communication and sensing (JCAS/ISAC), specifically focusing on OFDM-based agile beamforming, CDMA-coded nested arrays, frequency/time-modulated arrays (FMA/TMA), and direct RF/mmWave modulation arrays. Comprehensive analysis and study demonstrate architecture-specific trade-offs in spectral efficiency, AoA accuracy, latency, beam agility, and susceptibility to hardware imperfections. OFDM-based agile beamforming demonstrates precise AoA estimation and high beam agility, though latency is increased due to sub-symbol beam switching. CDMA architecture effectively enables simultaneous multibeam operation, but their performance critically depends on code orthogonality, length and baseband bandwidth. FMA and TMA efficiently encode spatial information into frequency and time domains respectively, achieving excellent angular resolution and rapid beamforming, however, requires precise time modulations and additional element-level multipliers for frequency synchronization. Direct RF modulation arrays achieve high spectral efficiency and scalability but require high speed and wideband circuit-level designs to manage impedance stability and switching-induced non-idealities. Consequently, selecting optimal architecture requires careful consideration of the target JCAS/ISAC application, balancing trade-offs among key system metrics, such as accuracy, latency, robustness, and energy efficiency, as well as circuit- and hardware-level constraints, to achieve next-generation JCAS/ISAC systems.

## Acknowledgement

The authors would like to thank Dr. Alberto Valdes-Garcia for his valuable feedback and Prof. Jane Gu from Georgia Tech for her invitation.

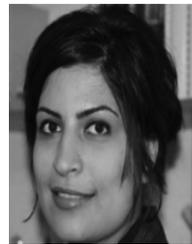

**Najme Ebrahimi** (Member, IEEE) received the B.S. (Hons.) from Shahid Beheshti University, the M.S. (Hons.) from Amirkabir University of Technology, and the Ph.D. from UC San Diego (2017). She was a Postdoctoral Fellow at the University of Michigan (2017–2020) and is currently an Assistant Professor at Northeastern University, previously at the University of Florida. Her research spans RF, mm-wave, THz ICs, wireless communications, IoT, and sensing. She received the DARPA YFA (2021), DARPA Director's Fellowship (2023), and serves on multiple IEEE CICC, IMS, BCICTS technical and steering committees.

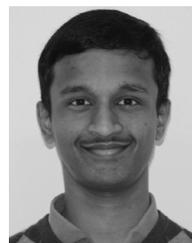

**Arun Paidimarri** (Senior Member, IEEE) received the B.Tech. from IIT Bombay (2009) and the S.M./Ph.D. from MIT (2011/2015). He is a Senior Research Scientist at IBM T. J. Watson and was an Adjunct Assistant Professor at Columbia University (2022). His research focuses on low-power wireless systems, mm-wave circuits, and software-defined phased arrays. He co-received Best Paper Awards at IEEE RFIC 2024, ICC 2013, and SmartCom 2019, holds three IBM Research Accomplishment Awards, and received the President of India Gold Medal (2009). He also won a Silver Medal at IChO 2005 and serves on IEEE TPCs.

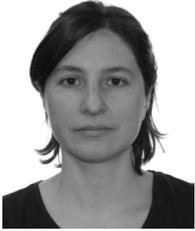

Alexandra Gallyas-Sanhueza (IBM T. J. Watson Research Center) received the B.Sc. in electrical engineering from Pontificia Universidad Católica de Chile, Santiago, Chile (2009-2015). She worked at Wilson Synchrotron Laboratory, Cornell University, Ithaca, NY (2017-2018). Alexandra received her M.Sc. and Ph.D. in electrical and computer engineering from Cornell University (2018-2021 and 2018-2024, respectively). During her PhD, she completed three summer internships, at the Analog Garage, Analog Devices (2021) and IBM T.J. Watson Research Center (2022 and 2023). She is currently a Research Scientist at the IBM T.J. Watson Research Center, Yorktown Heights, NY.

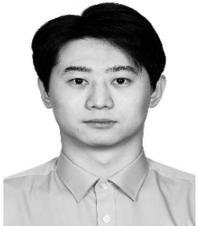

Yuan Ma (Graduate Student Member, IEEE) received his B.E. degree in Electronic Science and Technology from Beijing University of Technology (BJUT), Beijing, China, in 2023, and his M.S. degree in Electrical Engineering from Columbia University, New York, NY, USA, in 2025. He is currently working towards his Ph.D. degree in Electrical Engineering under the guidance of Prof. Ebrahimi at Northeastern University, Boston, MA. His research interest includes design of analog, RF, millimeter-wave and THz circuit and system.

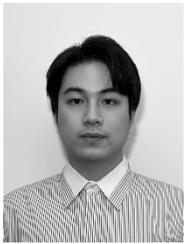

Haoling Li (Graduate Student Member, IEEE) received his B.E. degree in Electrical Engineering and Automation from Nanjing Tech University, Nanjing, China, in 2020, followed by an M.S. degree in Electrical and Computer Engineering from Northeastern University, Boston, MA, USA, in 2023. Currently, he is pursuing a Ph.D. in Electrical Engineering at Northeastern University, Boston, MA, under the supervision of Dr. Ebrahimi in her research lab. His research focuses on RF, millimeter-wave, and terahertz circuits and systems.

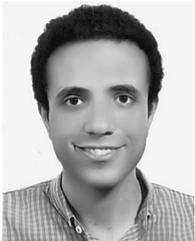

Basem Abdelaziz Abdelmagid (Graduate Student Member, IEEE) received the B.Sc. and M.Sc. degrees from the Electronics and Electrical Communications Engineering Department, Cairo University, Giza, Egypt, in 2018 and 2021, respectively. He is currently pursuing the Ph.D. degree with the Department of Information Technology and Electrical Engineering (D-ITET), Swiss Federal Institute of Technology Zürich (ETH Zürich), Zürich, Switzerland., His research interests include mm-Wave/sub-THz integrated circuits and systems, and power management integrated circuits for energy harvesting applications.

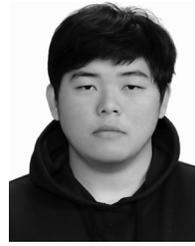

Tzu-Yuan Huang (Graduate Student Member, IEEE) received the B.S. degree in electrical and computer engineering from National Chiao Tung University (NCTU), Hsinchu, Taiwan, in 2012, and the M.S. degree from the Graduate Institute of Communication Engineering, National Taiwan University (NTU), Taipei, Taiwan, in 2015. He received his Ph.D. degree in electrical engineering at the Georgia Institute of Technology, Atlanta, GA, USA. His current research interests include RF/millimeter-wave (mm-wave) integrated circuits and systems. He is currently with ARGUS SPACE AG. He also serves on the Technical Program Committee (TPC) for IMS (2025–present).

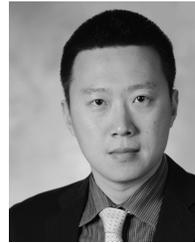

Hua Wang (Fellow, IEEE) received the M.S. and Ph.D. in Electrical Engineering from Caltech in 2007 and 2009. He was a faculty member at Georgia Tech (2012–2021), where he founded the CCS and directed the GEMS Lab. Since 2021, he has been a Full Professor and Chair of Electronics at ETH Zürich, leading the IDEAS Group. His research focuses on analog, mixed-signal, RF, and mm-wave ICs for communications, sensing, and bioelectronics. He has received the DARPA Director's Fellowship (2020), DARPA YFA (2018), NSF CAREER (2015), IEEE MTT-S Outstanding Young Engineer (2017), and multiple best paper awards.